# Critical cooling rates for amorphous-to-ordered complexion transitions in Cu-rich nanocrystalline alloys

Charlette M. Grigorian [a], Timothy J. Rupert [a, b, c, *]

[a] Department of Chemical and Biomolecular Engineering, University of California, Irvine, California 92697
[b] Department of Materials Science and Engineering, University of California, Irvine, California 92697
[c] Department of Mechanical and Aerospace Engineering, University of California, Irvine, California 92697
[*] Email address: trupert@uci.edu

**Abstract**

Amorphous complexions in nanocrystalline metals have the potential to improve mechanical properties and radiation tolerance, as well as resistance to grain growth. In this study, the stability of amorphous complexions in binary and ternary Cu-based alloys is investigated by observing the effect of cooling rate from high temperature on the occurrence of amorphous-to-ordered complexion transitions. Bulk Cu-Zr and Cu-Zr-Hf alloy samples were annealed to induce boundary premelting and then quenched through a procedure that induces a gradient of local cooling rate through the sample height. Amorphous complexion thickness distributions were found to be invariant to local cooling rate in the Cu-Zr-Hf alloy, demonstrating enhanced stability of the amorphous complexion structure compared to the Cu-Zr alloy, which had thinner amorphous complexions in the regions that were slowly cooled. The experimental results are used to construct time-temperature-transformation diagrams for the amorphous-to-ordered complexion transition in both the binary and ternary alloys, enabling a deeper understanding of the influence of cooling rate and grain boundary chemistry on complexion transitions. The critical cooling rate necessary to avoid complexion transitions in the ternary alloy is found to be at least three orders of magnitude slower than that for the binary alloy.

# 1. Introduction

The concept of *grain boundary complexions*, also termed *grain boundary phases* by some researchers [1-4], has helped elucidate previously unexplained phenomena in materials science. These grain boundary complexions are phase-like interfacial structures which exist in thermodynamic equilibrium with their abutting phases [5, 6]. Dillon et al. [5] classified six different types of grain boundary complexions found in undoped and doped alumina systems based on their thickness parallel to the grain boundary normal. However, we note that complexion thickness is only one of many possible descriptors, as the other various degrees of freedom such as structural and chemical disorder can be used to describe the complexion state. Transitions between these different complexions are also possible [7, 8], enabling control of the dominant complexion type in a microstructure, which in turn determines material properties [9-11]. For example, the fracture toughness of pure alumina with ordered complexions is greater than alumina doped with yttria or silica, which contained disordered complexions associated with embrittlement of the grain boundaries in that material [12]. Disordered complexion types are of particular interest given the unique properties they can impart to a material. Amorphous complexions form and are thermodynamically stable at elevated temperatures when a premelting transition occurs at grain boundaries [13, 14]. In contrast, ordered boundary structures are preferred at low temperatures. While amorphous complexions were first discovered in ceramic systems [15-19], their formation in a variety of binary metallic alloys has been demonstrated recently, including Mo-Ni [20], W-Ni [21], Ni-W [22], Cu-Bi [23], Ni-Zr, Cu-Zr, and Cu-Hf [24]. Schuler and Rupert [24] developed rules to predict complexion evolution, highlighting the tendency of dopant elements with a positive enthalpy of segregation and negative enthalpy of mixing to form amorphous grain boundary complexions. This is analogous to bulk metallic glass (BMG) formation, which generally requires a negative heat of mixing and an atomic size mismatch greater than 12% (atomic size is also important for determination of segregation enthalpy) [25, 26], although BMGs also have the added requirement of three or more elements. While amorphous complexions do not require such compositional complexity, this concept will be revisited shortly to discuss whether it would be beneficial.

The stability of amorphous grain boundary phases is of particular interest, as well as the processing necessary to lock in the high temperature grain boundary structure. To maintain amorphous complexions at ambient conditions and their associated benefits, a material must be quenched such that the high temperature grain boundary structure is retained. Thick amorphous grain boundary complexions are of particular interest in nanocrystalline metals due to their ability to improve thermal stability against grain growth by reducing the grain boundary energy and providing a heavily doped boundary that kinetically slows migration [22, 27, 28]. Activated sintering is also observed in materials containing amorphous intergranular films (AIFs), due to the enhanced diffusion which takes place with this type of grain boundary complexion [29-32]. Prior work by Donaldson and Rupert [32] showed that amorphous complexions can be used to create fully dense bulk nanocrystalline samples due to activated sintering, evidenced in this case by a large increase in sample density, arising at the temperatures where amorphous complexions form. Amorphous complexions have also been shown to improve resistance to radiation damage by limiting the size and number of defect clusters formed during irradiation [33, 34]. It is also desirable to have thick amorphous complexions within a material's microstructure due to their ability to enhance mechanical properties, as the ability of an amorphous complexion to absorb dislocations, to consequently resist cracking at the grain boundary and increase a nanocrystalline materials toughness, increases with increasing amorphous complexion thickness [35].

Khalajhedayati et al. [36] demonstrated that such a prediction is true for the Cu-Zr system, showing that strain-to-failure of a nanocrystalline alloys with amorphous complexions was much higher than that of the same alloy with ordered interfaces.  In recent years, a number of other groups have confirmed the toughening effect of amorphous interfacial films.  Wu et al. [37] reported on an Al alloy comprised of nanograins that were surrounded by nanometer-scale metallic glass shells, with micropillar compression testing of this composite structure showing a strength and homogeneous plastic strain that were much larger than either the nanocrystalline alloy or the metallic glass by themselves.  In a separate paper, Wu et al. [38] developed a material which contained a nanocrystalline high entropy alloy phase that was wrapped by nano-sized amorphous regions, finding that the material exhibited a ductile response with a near-theoretical strength.  Finally, Yang et al. [39] showed that disordered interfaces can enable ductile ordered superlattice alloys by preventing brittle intergranular fracture.  Hence, the presence of amorphous complexions is beneficial for a number of properties and behaviors, with thicker complexions generally being better.

While the vast majority of studies on amorphous complexions focus on their formation in binary alloys, there are demonstrated benefits in the literature associated with incorporating additional dopant elements.  Zhou et al. [40] found that alloys with complex grain boundary compositions, containing 5 or more elements in this case, were able to retain nanocrystalline crystallite sizes to higher temperatures than their pure Ni or Ni-rich binary counterparts.  These authors hypothesized that the greatly enhanced thermal stability of these alloys is due to what they termed "high-entropy grain boundary complexions," allowing for greater reduction of GB energy at higher temperatures [41].  In a different study, Zhou and Luo [42] showed that multicomponent alloys enable activated sintering due to the increased disorder at the interface caused by co-segregation of dopant atoms. Grigorian and Rupert [28] demonstrated that the use of an additional dopant element forms thicker amorphous complexions in Cu-Zr-Hf alloys compared to binary Cu-Zr and Cu-Hf alloys with similar compositions. In that study, alloys in powder form were quenched in a water bath while under vacuum, with the amorphous complexion population captured during the quench then measured. Thicker amorphous complexions were reported in the Cu-Zr-Hf alloys than the two binary alloys investigated, implying that the equilibrium complexion thickness is affected by the chemical complexity at the grain boundary. However, the samples were either "quenched" or "slowly cooled" in Ref. [28], with the temperature history in neither process being quantifiable. The importance of cooling rate on the transition of amorphous complexions back to an ordered state has been unexplored to date. Segregation of multiple dopant elements is expected to also improve the stability of amorphous complexions by enhancing their resistance to crystal nucleation in a manner similar to BMGs. BMGs are similarly metastable at room temperature and therefore a parallel may be drawn to strategies from BMG processing to identify strategies for stabilize amorphous complexions. Increasing the number of constituent elements used to form BMGs has been proven to improve their glass forming ability, typically measured in terms of the cooling rate required to maintain an amorphous structure and therefore miss the crystalline phase region on a time-temperature-transformation (TTT) diagram [43-47]. Men et al. [48] demonstrated this principle in CuZr-based BMGs, where the addition of 7.5 at.% Ti increased the critical casting diameter from 2 mm to 5 mm, indicating an improvement in the glass forming ability. Amorphous complexion formation has been investigated in a variety of binary and multicomponent alloys. However, the resistance to amorphous-to-ordered complexion transitions has not yet been compared across chemistries, nor have complexion TTT diagrams been created for this transition. Therefore, a thorough understanding of this transformation and the dependence on grain boundary chemistry is timely.

In this work, we investigate Cu-Zr and Cu-Zr-Hf alloys to decipher the time-temperature-transformation behavior of amorphous complexions during cooling, providing the first report on the kinetics and phase transformation fundamentals associated with such a transition. Fully dense bulk samples of binary and ternary nanocrystalline alloys with matching global dopant concentration were annealed at high temperature (~98% of the melting temperature, $T_m$) to form thick amorphous complexions. The samples were cooled in a controlled fashion from one side only to induce variable quenching/cooling rates throughout the material that could then be used to interpret variations in complexion structure. Amorphous complexion thickness distributions were measured at various positions within the two alloys to investigate their relative stability as evidenced by variation or lack thereof in the thickness distribution as cooling rate changes. Within the range of quench rates accessible with this experimental setup, a direct relationship between amorphous complexion thickness and cooling rate is established for the bulk Cu-Zr alloy. In contrast, the bulk Cu-Zr-Hf alloy retains amorphous complexions with a thickness population that is invariant to cooling rate, suggesting that the transformation back to ordered boundaries is completely avoided for the rates probed here. For both alloys, the fastest cooled regions contained amorphous complexions that were notably thicker than prior reports from quenched powders [28]. The information regarding the relationship between complexion thickness and cooling rate is then used to develop preliminary time-temperature-transformation diagrams for both alloys. The results of this study demonstrate that alloys with multiple dopant elements form amorphous complexions which are more resistant to transforming back to the ordered complexion state, with the critical cooling rate necessary to avoid transformation being more than three orders of magnitude slower than that of the binary counterpart. From a practical standpoint, these findings can be used to enable the formation of amorphous complexions in larger bulk metallic samples, which, with their unique properties, can impart advantages to materials in structural, refractory, or nuclear applications.

## 2. Materials and Methods

Cu-3 at.% Zr and Cu-2 at.% Zr-1 at.% Hf alloys were produced from powders of elemental Cu (Alfa Aesar, 99.99%, -170 + 400 mesh), Zr (Micron Metals, 99.7%, -50 mesh), and Hf (Alfa Aesar, 99.8%, -100 mesh). The total global dopant concentration was chosen to be constant at 3 at.% to best compare the complexion transition behavior between the binary and ternary alloys. These samples are simply referred to from here on as Cu-Zr and Cu-Zr-Hf, respectively. The starting powders were ball milled for 10 h in a SPEX SamplePrep 8000M high-energy ball mill using a hardened steel vial and milling media. A ball-to-powder weight ratio of 10:1 was used with 1 wt.% stearic acid added to the powders prior to milling as a process control agent, in order to prevent excessive cold welding of the powders. The individual alloy compositions were confirmed using energy dispersive X-ray spectroscopy (EDS) in a Tescan GAIA3 scanning electron microscope (SEM) equipped with a 150 mm$^2$ Si drift detector. Following the mechanical alloying process, phase identification and crystallite size determination for the as-milled powders were performed in a Rigaku SmartLab X-ray diffractometer (XRD). The resulting Cu-Zr and Cu-Zr-Hf alloys had crystallite sizes of 25 and 29 nm for the face centered cubic (FCC) phases, respectively, measured via Rietveld refinement of XRD patterns using MAUD analysis software [49]. The XRD results for crystallite size and phase percent for the FCC and carbide phases in the samples are compiled in Table 1.

The alloy powders were consolidated to form bulk samples using an MTI Corporation OTF-1200X-VHP4 hot press. The powders were first cold pressed under 25 MPa for 10 minutes in a graphite die. The resulting green bodies were then hot pressed under 50 MPa for 10 hours at a temperature of 900 °C, allowing for full densification of the samples, as well as the segregation of dopant elements to the grain boundaries. XRD measurements after consolidation show that some minor grain growth of the FCC phase occurred during the hot pressing, with final crystallite sizes of ~60 nm being measured for both alloys (Table 1). According to prior work by Donaldson and Rupert investigating the densification behavior of Cu-4Zr alloys with amorphous complexions, the relative density of bulk samples formed under these particular pressing conditions was found to be 99.7% [32]. It is important to note that while amorphous complexions do form at grain boundaries during sintering, they are not retained after this process due to the slow cooling of the bulk piece back to room temperature while in the hot press, which allows time for the glassy interfacial complexion to transition back to an ordered structure. The resulting cylindrical bulk samples had a height of 6.47 mm and a radius of 5.34 mm, and were cut in half lengthwise to form a semi-cylinder, as shown in Figure 1(a), prior to subsequent annealing in a tube furnace at 950 °C for 10 minutes to re-form the amorphous complexions. The samples are then removed from the furnace as quickly as possible and placed on an Al heat sink that is cooled by a liquid nitrogen reserve (Figure 1(a)). This allows for the sample to be rapidly cooled, while also inducing a gradient of cooling rates as a function of distance from the surface of the heat sink. The surface directly in contact with the aluminum (bottom) will be very quickly quenched compared to the interior (middle) and opposite side of the sample (top). A thin oxide film formed at the surface of the sample upon transfer from the furnace to the Al block was removed prior to further processing. The semi-cylindrical samples were then cut in half once more to expose the center of the sample for characterization in order to investigate the effect of varying the cooling rate on the local grain boundary structure.

The cooling rate gradient induced and its connection to the measured complexion population are expected to reveal the stability of the amorphous complexions against transformation to thinner complexions or ordered grain boundaries. The population of amorphous complexions which form in a particular alloy is associated with the alloy composition and the annealing temperature. However, the high temperature complexion population may differ from what is observed at room temperature, as some amorphous boundaries could transform back to an ordered grain boundary structure if the region is not quenched quickly enough. This is equivalent to what is observed regarding the nucleation of crystalline regions in BMGs when cooled below the critical cooling rate necessary to retain a completely amorphous microstructure [50, 51]. Complexion transformations may be even more sensitive to variations in cooling rate due to the very nanometer length scales at which complexion transitions occur. Therefore, we expect that the section of the sample which is quenched the fastest and is in direct contact with the Al heat sink would be closest to the high temperature microstructure, and will be henceforth referred to as the *fast quench* surface. The center of the sample is referred to as the *medium quenched* region, while the top of the sample farthest from the heat sink is referred to as the *slow quenched* region.

Transmission electron microscopy (TEM) specimens were taken from the Cu-Zr and Cu-Zr-Hf samples at the fast, medium, and slow quenched sections of the sample using the focused ion beam (FIB) lift-out method with a Tescan GAIA3 SEM/FIB. A final 5 kV polish was used on all samples to remove any excess ion beam damage from earlier preparation steps. Microstructural characterization, including bright field TEM (BF TEM), selected area electron diffraction (SAED), STEM-EDS lines scans of grain boundary composition, high resolution TEM (HRTEM) of the grain boundary structure, measurement of AIF thickness, and high-angle annular dark field STEM (HAADF) were performed using a JEOL JEM-2800 TEM. The thicknesses of at least 20-40 amorphous complexions from the fast, medium, and slow quenched regions of both the Cu-Zr and Cu-Zr-Hf samples were measured by orienting the grain boundaries in an edge-on condition, which was confirmed by ensuring that there is no variation in the thickness of the complexion when viewed in under-focused and over-focused imaging conditions.

## 3. Results and Discussion

*3.1 Simulation of Temperature Distribution within the Bulk Samples*

To provide insight into the cooling rate variation within the sample, a COMSOL Multiphysics simulation of the cooling process was performed, as shown in Figure 1(b). Table 2 compiles the sample properties and boundary conditions used for this model of the quenching process. A metal semi-cylinder (thermal conductivity of pure Cu is used for the consideration of heat transfer properties) with the aforementioned sample dimensions was constructed. To replicate the experimental cooling conditions, a temperature of 77 K is assigned to the bottom surface of the sample where it contacts the Al cooling block. The cooling block is much larger than the nanocrystalline pieces and is partially submerged in a large reservoir of liquid nitrogen reservoir, so the Al block is assumed to remain at 77 K during cooling of the specimen. Conductive heat transfer occurs from this bottom surface, while the other sample surfaces are cooled by radiation and convection into the air at room temperature.

Figure 1(b) shows the temperature profile within a bulk piece 0.05 s after contact with the cooling block. The blue dot represents the fast quench side of the sample which is directly in contact with the Al block, while the orange and red dots represent the medium (middle) and slow (top) quenched regions of the sample, respectively. It is also important to note that temperature and cooling rate data for the fast and slow quenched regions of the sample, as well as the TEM specimen extraction described later, were obtained at a distance of ~100 μm from the surface of the sample in order to avoid the effects of a thin oxide layer which forms on the surface when the hot piece is exposed to air during transfer from the furnace. The temperature of each of these three regions is plotted as a function of time ($t$) in Figure 2(a), with $t = 0$ representing the initial contact with the cooling block.

The quenching at different locations can be quantified in a number of ways, with the first being the time required for sample position to reach a given temperature. This would be most useful if specific target temperatures such as the glass transition temperature for the amorphous complexion or the temperature associated with the "nose" (i.e., the temperature where nucleation and growth competition is optimized) of a transition to an ordered complexion from an interfacial TTT diagram were known. These temperatures are not known currently, although some approximate values can be hypothesized. For example, a prior study by Khalajhedayati and Rupert suggests that the transition to an amorphous complexion for Cu-3Zr occurs somewhere between 750 °C and 850 °C, in the range of roughly 82% to 90% of the solidus temperature for this alloy [27]. This is in agreement with Luo et al. [52], who have predicted that grain boundary premelting transitions can begin at temperatures as low as 60-85% of the melting temperature. The upper bound of relevant temperatures would therefore be roughly $0.85T_m$ or 785 °C, the highest temperature below which transitions from an amorphous to an ordered complexion would be expected. The lower bound of this range at $0.6T_m$ or 474 °C represents a possible intermediate temperature to consider, as ordered complexions would be preferred while there is still enough thermal energy for rapid diffusion. A possible lower bound would be where diffusion has slowed significantly, which we take as a temperature of $0.4T_m$, or 225 °C, to match the description of rough bound for a "high temperature" when considering other diffusion-based processes such as creep. These temperatures are not meant to be exact values of great importance, but rather to give a rough sense of the important cooling times. Figure 2(b) displays the time required for each location to reach each of these three temperatures. The entire sample cools to a temperature of $0.4T_m$ in less than 0.2 s (reminding the reader that this is not accounting for the short amount of time required to quickly transfer the sample from the furnace to the heat sink).

To gain a better understanding of the quenching at each position, the derivative of the cooling curves or the cooling rate, $\dot{T}$, is shown in Figure 2(c), while Figure 2(d) shows the maximum cooling rate, $\dot{T}_{max}$, for each sample position. We note that the maximum cooling rate for the fast quench location, shown in blue, is much faster than the other two locations, so this peak is out of the y-axis range in Figure 2(c). The results reported in Figure 2 confirm that the chosen method of controlled quenching achieves a variety of cooling rates at different sample positions, with $\dot{T}_{max}$ ranging from ~5900 K/s in the slow quench region to $1.03 \times 10^6$ K/s in the fast quench region, spanning multiple orders of magnitude. These cooling rates are on par with or exceed those necessary to form bulk metallic glasses in Cu-Zr and Cu-Hf alloys via processing methods such as melt spinning [53-55]. The significant differences in the cooling rates between sample positions allows for the investigation of the impact of quench rate for the retention of AIFs in both the binary and ternary nanocrystalline alloys.

*3.2 Characterization of Microstructure, Grain Boundary Chemistry, and Interfacial Structure*

Detailed characterization of the microstructure at each of the fast, medium, and slow quench positions for the Cu-Zr and Cu-Zr-Hf samples was next performed in order to probe crystallite size and grain boundary segregation state. HAADF micrographs of representative microstructures of each of the three investigated sample positions in the binary and ternary samples are shown in Figure 3(a). The average crystallite sizes for each sample location, plotted in Figure 3(b), are nearly identical, confirming that the variations in cooling/quenching rate do not induce a gradient in crystallite size or provoke abnormal grain growth. The average crystallite sizes were found to be between 67 nm and 73 nm in all cases, showing very small variations within the expected noise in the measurement and being comparable to the XRD measurements. The similar crystallite sizes mean that the volume of grain boundary material is also nearly the same for all samples investigated in this study. In Figure 3(a), the presence of second phase precipitates throughout the microstructures of both the Cu-Zr and Cu-Zr-Hf samples is found to be minimal, confirming XRD measurements in Table 1 of ~1 vol.% of carbides in the samples, and therefore should not significantly influence the amount of dopant that is available to segregate to grain boundaries. More detailed characterization of the limited ZrC and HfC second phases that form in Cu-Zr and Cu-Zr-Hf microstructures can be found in our prior study of these alloys [28]. Bright contrast at the grain boundaries is observed for both samples and at all positions, indicating locations where Zr and/or Hf dopant atoms have segregated during either the powder consolidation and/or annealing steps. To study grain boundary segregation more explicitly, STEM-EDS line scans were performed on multiple boundaries in each consolidated alloy, with representative examples presented in Figure 4. For both the Cu-Zr and Cu-Zr-Hf alloys, the alloying elements were found to be strongly segregated to the grain boundary region, with high EDS intensity measured at the interface and low values near zero measured in the grain interior. Heavily doped boundaries are likely sites for the formation of amorphous complexions, as increased local composition above a critical threshold value can lead to the transition to an amorphous structure at high temperatures [56]. The fact that the samples have the same global dopant concentration, crystallite size, grain boundary volume fraction, segregation state, and carbide content allows for a direct comparison of complexion populations between samples, with only the complexity of the grain boundary chemistry (binary versus ternary) and quench rate (fast, medium, or slow) changing.

HRTEM of the grain boundaries was next performed. We focused on identifying interfaces which had maintained the amorphous complexion structure and then the subsequent measurement of the thicknesses of these structures. Micrographs of representative amorphous complexions found in the Cu-Zr alloy are shown in Figure 5. Very thick amorphous complexions are found regardless of the maximum cooling rate at each sample position. However, we emphasize that there are boundary-to-boundary variations in amorphous complexion thickness, which has been shown previously to be related to variations in dopant concentration that depend on boundary character [56]. It is also important to note that not every grain boundary within the microstructure is an amorphous complexion, as some doped but ordered interfaces are still found. We observe here that amorphous complexions are more likely to form at grain boundaries where high levels of dopant segregation are observed, such as those of high contrast noted in Figure 3(a), a finding that is consistent with prior work on these alloys [6, 57].

Figure 6(a) shows a micrograph of an interesting grain found in the fast quench region of the Cu-Zr sample which had the majority of its boundaries in an amorphous state. An example of an HRTEM image of the boundary along the top of the grain is shown in Figure 6(b), confirming that it is an amorphous complexion. The overlay of red and blue dashed lines on Figure 6(c) represent ordered and amorphous grain boundaries, respectively. This grain demonstrates the efficacy of rapid quenching for freezing in many thick amorphous complexions within the microstructure. It is important to note that measurements of complexion thickness, presented later in the work, are not taken from low magnification images such as the one shown in Figure 6. Measurements of amorphous complexion thickness are obtained at very high magnification and exclusively from those boundaries in an edge-on condition, verified by ensuring there is no variation in thickness when viewing the amorphous boundary in under- and over-focused imaging conditions. In addition, grain boundaries in nanocrystalline alloys often have a noticeable curvature on the length scales associated with TEM investigation, meaning that a given interface can have a changing grain boundary plane normal within a section being investigated by HRTEM. The thickness of an amorphous complexion depends on the grain boundary character, meaning that one must be careful and follow a consistent procedure for such measurements. To this point. amorphous complexion thickness measurements are taken from the thinnest observed amorphous region in order to ensure consistency in the reported measurements, as shown in Figure 7. For the amorphous complexion shown here, a gradient in film thickness is observed as the boundary curves, with the lowest thickness (4.7 nm in this case) being tabulated as the measurement for each inspected grain boundary region.

Representative micrographs of amorphous complexions found in the Cu-Zr-Hf alloy from the fast, medium, and slow quenched regions are shown in Figure 8. Again, thick amorphous complexions are found in all locations. The rapid cooling rates during the bulk sample quench against the Al heat sink apparently allow amorphous complexions to be stabilized for both alloy systems. In order to compare the effect of these varying cooling rates on amorphous complexion populations in the two bulk samples, 20-45 amorphous complexions were isolated and the thickness measured in an edge-on condition from the fast, medium, and slow quench regions of the binary and ternary samples. Cumulative distribution functions of amorphous complexion thickness at each of the three positions in the Cu-Zr sample are shown in Figure 9(a). These complexion thickness distributions reveal an obvious dependence of the ability to retain thick amorphous complexions on the local cooling rate. The fast quench side of the sample sustains by far the thickest amorphous complexions (average thickness of 2.9 nm, with a number of very thick amorphous films above 4 nm), while the thinnest ones are found on the slow quench side of the sample (average thickness of 1.3 nm). To put the amorphous complexion thickness populations into perspective, we compare these results with the amorphous complexion thickness results from a prior study where a Cu-4Zr powder sample encapsulated under vacuum in a quartz tube was quenched by dropping into a water bucket from the same annealing temperature [28]. While the sample was quickly put into the quenching medium in [28], there was likely some delay in the reduction of the temperature of the powder particles. The distribution of amorphous complexion thicknesses measured in the quenched powder sample are also plotted in Figure 9(a), with an average thickness of 1.6 nm. The rapid quenching method used in this study allows for a larger number of thick amorphous complexions to be sustained in the Cu-3Zr microstructure, with roughly double the average complexion thickness compared to the Cu-4Zr powders quenched in a quartz vial. While amorphous complexions were achieved in all specimens through the careful selection of dopant elements which segregate to grain boundaries, the results of this study highlight the importance of processing conditions, in particular cooling rate, to capture thick amorphous complexions by preventing transitions to thinner amorphous films or ordered grain boundary phases altogether.

In contrast, the complexion thickness distributions for the Cu-Zr-Hf alloy shown in Figure 9(b) show no dependence on the cooling rate. This resistance to transforming to ordered grain boundaries under identical cooling profiles can be attributed to the resulting reduction in the energy for the amorphous phase with a more complex chemistry, moving from two elements at the grain boundary to three, which further stabilizes the amorphous grain boundary phases that are able to form [5, 58, 59]. The addition of three or more alloying elements is a widely used method of improving the stability of bulk metallic glasses, whereby increasing the number of constituent elements enables the formation of a highly dense, randomly packed amorphous structure while hindering the nucleation and growth of a crystalline phase due to the increased difficulty of long-range atomic rearrangements [26, 60]; the same materials design concept appears to also improve the stability of amorphous complexions. We also note that the amorphous complexion populations in the Cu-Zr-Hf sample are nearly identical to that of the fast quench region of the Cu-Zr sample. This similarity suggests that these amorphous complexion populations are likely representative of the high temperature state, while the other two populations in the binary alloy represent states where some of the amorphous boundaries have either become thinner or transformed back into ordered grain boundaries. While our prior study reported that thicker amorphous intergranular films were found in Cu-Zr-Hf than Cu-Zr due to the more complex grain boundary chemistry [28], the difference observed in that study may have been associated more with the slower cooling rate associated with quenching the powders than the difference in compositional complexity. We hypothesize that the amorphous complexion population in the binary alloy in [28] became thinner more quickly than the ternary alloy during the cooling step, rather than there being a difference in the complexion structure at high temperature.

*3.3 Time-Temperature-Transformation Diagrams for Amorphous-to-Ordered Complexion Transitions*

The equilibrium structure and thickness of a grain boundary complexion is largely determined by the temperature and grain boundary composition, so it is expected that amorphous complexions will thin or even undergo a transition to the ordered structure as the temperature of the material is reduced. However, analogous to bulk phase transitions, this complexion transition would not happen immediately upon cooling due of the time required for an ordered complexion to nucleate and grow along the boundary. We hypothesize that the addition of Hf to a Cu-Zr alloy adds additional chemical complexity to the grain boundary region, making the transition to an ordered grain boundary more difficult and therefore making the complexion more resistant to transformation. The significant differences in the retained amorphous complexion populations in the Cu-Zr for different cooling rates motivate a more in-depth analysis of the amorphous-to-ordered complexion transformation in these two alloys.

The relative stability of the amorphous complexions formed in Cu-Zr and Cu-Zr-Hf can be understood using a schematic TTT diagram such as the one shown in Figure 10(a). The blue, yellow, and red arrows represent the fast, medium, and slow quench cooling paths, respectively. Conditions that provoke transition to thinner amorphous or ordered grain boundary complexions in the binary Cu-Zr alloy are represented by the shaded green region, while the more stable ternary Cu-Zr-Hf alloy transformation region appears as a shaded purple region. With our hypothesis that the complexion population at the fast quench region of the Cu-Zr sample is representative of the high temperature grain boundary structure, the transformation region is drawn to the right of the fast cooling curve. This indicates that this region is cooled at a rate which is greater than the critical cooling rate ($R_c$) necessary to retain the population of thick, premelted boundaries formed during annealing. However, at regions quenched slower than $R_c$, represented by the slow and medium quench curves, amorphous complexions undergo a complexion transition, resulting in thinner complexion populations. In contrast, the nose of the curve representing the transition to thinner amorphous or ordered complexions in the ternary Cu-Zr-Hf alloy is shifted to the right on Figure 10(a), indicating that the premelted grain boundary containing three elements is more stable against crystallization for a longer period of time as compared to the binary alloy. Therefore, $R_c$ is reduced and the cooling rates of the fast, medium, and slow quenched cooling rates are sufficient to avoid the complexion transition for Cu-Zr-Hf, as evidenced by the identical amorphous grain boundary thickness distributions that are observed.

We note that the schematic view in Figure 10(a) does not distinguish between the different complexion transitions that can result during cooling. An amorphous complexion of a given thickness at the annealing temperature would want to take on a smaller thickness at a lower temperature, but this transition would take time to occur. Therefore, a more complete TTT diagram would have many such transition regions defined, one each for each amorphous complexion thickness and another for the final reversion to an ordered structure. Such a representation would be unwieldy but also only hold exactly true for one given boundary. Another interface with a different segregation composition and a different crystallography would have these transition regions positioned at slightly different locations [6, 56, 61]. Since we are looking at the behavior of a larger population of amorphous interfacial films, we simplify this view to have all transitions to either thinner amorphous films or to an ordered structure outlined by a single transition region for each alloy. A more complete description of the connection between the thickness of a single amorphous complexion and temperature is available in a number of prior studies, expressed as grain boundary phase diagrams [42, 62, 63].

The extent to which amorphous complexions in the medium and slow quench regions of the Cu-Zr samples have transformed to thinner or ordered boundaries during cooling can be estimated by comparing the amorphous complexion thickness distributions between the different quench rates. Because the complexion population in the fast quench region of the binary alloy matches the invariant amorphous complexion populations found in the ternary alloy, we make the assumption that it is representative of the equilibrium complexion thickness population at high temperatures (i.e., no thick amorphous-to-thin amorphous or amorphous-to-ordered complexion transitions have occurred). To represent the transition of a fraction of the original amorphous complexions, the thickest measurements are removed one by one from the fast quench population and the remaining populations are replotted as cumulative distribution functions in Figure 10(b). In this figure, the dashed grey lines represent subsets of the fast quench complexion thickness data, where the bottom X% of the high temperature or fast quenched distribution is plotted, meaning that 100-X% of the amorphous boundaries have become thinner, either remaining amorphous but with reduced thickness or transforming to an ordered structure. Using these curves, we can estimate the fraction of complexions in the medium and slow quenched regions which have transformed. The medium quench region is best fit by X = 74%, implying that 26% of the original amorphous complexions have transformed. Similarly, the slow quench region is best fit by X = 37%, implying that 63% of the original complexions have transformed. We note that this analysis makes no specific claims about the behavior of any given boundary or implies any specific transition, other than the general shift of the statistical thickness distribution to the left as amorphous complexion thickness in general decreases. Figure 10(c) reports the fraction of amorphous complexions in the Cu-Zr alloy that have transformed to either a thinner amorphous or ordered complexion plotted as a function of the local maximum cooling rate, showing that the number of amorphous complexions which transform increases as the cooling rate decreases.

A quantitative comparison of $R_c$ for the transformation of thick, disordered to thinner or complexions boundaries in the Cu-Zr and Cu-Zr-Hf alloys can be demonstrated with TTT curves analogous to those developed for bulk metallic glasses. In order to construct these curves, we again must that the population of complexions in the fast quench regions are representative of the high temperature grain boundary structure. In addition, it is assumed that this region has cooled at a rate only just sufficient to avoid the complexion transition, and therefore the cooling path of the fast quench region narrowly misses the nose of the TTT curve for the binary alloy. TTT curves constructed for both the Cu-Zr and Cu-Zr-Hf alloy are shown in Figure 11. The temperature corresponding to the nose of the crystallization TTT curve for bulk metallic glasses has been previously shown to be roughly the average of the glass transition temperature and liquidus temperature, $T_l$, for the alloy [64, 65], indicated by the $0.5(T_g+T_l)$ line. The $T_l$ of a bulk metallic glass is analogous to the solidus temperature of the alloys in this study ($T_{solidus}$). $T_g$ is most closely analogous to the temperature at which ordered complexions transform to amorphous complexions ($T_{AIF}$), which is taken to be $0.85T_{solidus}$ to account for the upper temperature limit at which an amorphous-to-crystalline complexion transition can occur. [52]. The nose of the two TTT curves are placed at a temperature halfway between these limits, at 879 °C. The nose of the binary alloy transition region is positioned to be tangential to the fast quench cooling path, suggesting that while no complexion transition occurred, any reduction in quench rate would have led to a decrease in amorphous complexion thickness. This is an inherently conservative assumption. Although the cooling rates in our experiment are not constant, estimated constant cooling rate $R_c$ curve is plotted alongside actual cooling curve of the fast quench region to provide an estimate that a constant cooling rate of ~$6\times10^6$ K/s would be needed to avoid the complexion transition. The medium and slow quench regions of the bulk binary sample cool at considerably slower rates and traverse through the transformation region in the TTT curve, resulting in reconfiguration of the AIFs in these regions to either thinner AIFs or ordered complexions.

A similar TTT curve for the Cu-Zr-Hf alloy is drawn further to the right on this figure, indicating the observed resistance to the complexion transition. The fact that the quenching curves do not pass through the complexion transition region on the schematic illustration is justified by the complexion thickness distributions measured in the ternary alloy and the observation that these distributions appear to be insensitive to cooling rate within the boundaries probed here. To again make a conservative estimate, we again assume that the nose of the transition region would have begun at a time immediately after the time taken for the slow quench region of the sample to reach $T_{nose}$. Again overlaying a constant cooling curve, we calculate that the $R_c$ of the ternary alloy with be ~$1.8\times10^3$ K/s. Comparison of the estimated $R_c$ values therefore suggests that the ternary alloy can be cooled over 3000 times slower and still retain the complexion population that is representative of the high temperature equilibrium structure, demonstrating a significant improvement in the stability of thick amorphous complexions with increasing chemical complexity. It is important to note that the magnitude of reduction in $R_c$ is not strongly dependent on any of the assumptions made about the location of $T_{nose}$ for these TTT curves. For example, if a much lower $T_{nose}$ value of 225 °C (corresponding to $0.4T_{solidus}$ below which diffusion would be very slow) is alternatively chosen, $R_c$ in the Cu-Zr-Hf sample is still roughly three orders of magnitude lower than $R_c$ for the Cu-Zr sample.

The reduction of $R_c$ by a significant factor helps explains the observed difference in amorphous film thickness between Cu-Zr and Cu-Zr-Hf alloys in our prior investigation on powder specimens in Ref. [28]. In this prior work, fewer complexion transitions occurred, leading to an amorphous complexion population that was thicker and closer to the high temperature distribution.

In addition, the complexions found in the fast quench region of the binary sample and throughout the entire ternary sample are the thickest stable amorphous intergranular films reported to date. With an average thickness of 2.9 nm, the amorphous complexions are ~40% thicker than the complexion population observed in the ternary sample in Ref. [28]. With an eye toward future work and more rigorous confirmation of the high temperature structure, we expect that the most accurate evidence would come from in situ TEM heating experiments to directly observe equilibrium amorphous complexion thicknesses at high temperature.

The choice of elements with a positive enthalpy of mixing, negative enthalpy of segregation, a large atomic size mismatch, and the use of multiple constituent elements are all guidelines suggested to maximize the glass-forming ability (GFA) of metallic glasses. A variety of parameters, including but not limited to the inverse of $R_c$, are used to quantify the GFA of bulk metallic glasses. Improvements to GFA have been associated with the addition or substitution of dopant elements in these materials, analogous to the improvements in amorphous complexion stability reported in this study. For example, Yu et al. [66] demonstrated an improvement in GFA of a binary Cu-Zr alloy upon addition of a third and fourth element, evidenced by both a dramatic decrease in $R_c$ as well as an increase in the critical casting diameter. The selection of elements with a positive enthalpy of segregation, negative enthalpy of mixing, and multiple constituent elements has also been shown to promote the formation of amorphous complexions at grain boundaries in prior studies [24, 28]. However, the impact of the implementation of each of these criteria has not been proven to improve the stability of the amorphous complexions or shown to resist transition back to the ordered structure prior to this study. While the present results demonstrate a significant reduction in $R_c$ is achieved with the addition of a second dopant element, to achieve a ternary grain boundary composition, we note that this improved stability is achieved despite the fact that Zr and Hf have very similar atomic radii (Zr: 160 pm, Hf: 156 pm), with an atomic radius difference of 2.5%. This is consistent with several studies of bulk metallic glasses that have also shown that the replacement of a fraction of Zr with an appropriate amount of Hf can lead to a significant improvement in GFA [67-69]. One such example is a study of an Fe-based bulk metallic glass, where the GFA of the base alloy containing 8 at.% Zr was improved upon substituting 3 at.% Hf for 3 at.% Zr, with a reduction in glass transition temperature used to show the effect [69]. The results of this study confirm that increasing the compositional complexity of the grain boundary region helps to prevent the local atomic rearrangements necessary for the amorphous complexion to crystallize into an order boundary structure.

The ability to sustain thick amorphous complexions in a bulk material through the reduction of $R_c$ also holds significant important from a materials processing perspective. While the presence of amorphous complexions has been shown to improve the fracture toughness, strength, and ductility of nanocrystalline materials, these features must be sustained within the material's microstructure in order to impart these effects [35, 36, 70]. A simple practical observation from this work is that bulk samples were produced that contained thick amorphous complexions with a stable thickness distribution throughout the specimen, opening the doorway for larger nanocrystalline specimens with improved properties.

## 4. Conclusions

In this study, amorphous grain boundary complexions were investigated in binary and ternary Cu-rich alloys with an eye toward evaluating their relative stability against transformation back to an ordered structure when cooling from a high temperature. Bulk Cu-3Zr and Cu-2Zr-1Hf alloy specimens with nanocrystalline grain size were created by sintering mechanically alloyed powders, and were then annealed and quenched in a special manner in order to induce a cooling rate gradient. The fast, medium, and slow quench regions of both bulk samples were inspected using HRTEM to characterize and tabulate the thicknesses of the amorphous complexions. The imposition of a cooling gradient enabled the creation of a time-temperature-transformation diagram for the amorphous-to-ordered complexion transition, and allowed the critical cooling rates needed to lock in the high temperature, amorphous structure to be estimated. The following conclusions can be drawn:

- Amorphous complexions formed at all positions throughout both the binary and ternary bulk samples. The quenching method utilized in this study, in which bulk samples are placed on an Al heat sink immersed in liquid nitrogen, retained much thicker amorphous complexions compared to previous studies where powder samples of similar compositions were quenched into water baths inside of quartz vials. The increased amorphous film thickness in this study is attributed to achieving cooling rates greater than the critical cooling rate necessary to avoid amorphous complexions transforming back to thinner films or ordered boundaries. These samples demonstrate that the amorphous complexion materials design concept can be applied to bulk material forms, while past studies had only synthesized such features in microscale powders and films.

- Amorphous complexions in the Cu-Zr-Hf sample were significantly more stable against complexion transitions to thinner or ordered grain boundaries compared to the AIFs in Cu-Zr. On average, the complexions in the binary alloy were thinner at slower cooling rates, while the complexion thickness population for the ternary alloy was found to be invariant with cooling rate. This observation proves that the addition of a third element serves to stabilize amorphous grain boundary complexions, evidenced by a significant reduction of the critical cooling rate necessary to prevent the amorphous-to-ordered complexion transformation. The value of the critical cooling rate was found to be at least three orders of magnitude slower in the ternary alloy compared to the binary alloy.

The results of this study demonstrate the enhanced stability of thick amorphous complexions with ternary grain boundary chemistries, isolating increased chemical complexity as a key stabilization mechanism. This finding makes nanocrystalline alloys with amorphous complexions accessible by a greater range of conditions and enables applications that might benefit from the unique properties of amorphous complexions, such as enhanced thermal stability, fracture toughness, and resistance to radiation damage. This study also establishes that complexion TTT diagrams are valuable tools for understanding complexion transitions and the stability of complexion populations. The results of this study motivate further experimental and computational work to construct TTT diagrams for a variety of complexion transitions, as well as the investigation of the stability of amorphous complexions which form in more compositionally complex alloys.

**Acknowledgements**

This study was supported by the U.S. Department of Energy, Office of Basic Energy Sciences, Materials Science and Engineering Division under Award No. DE-SC0021224. SEM, FIB, TEM, and XRD work was performed at the UC Irvine Materials Research Institute (IMRI) using instrumentation funded in part by the National Science Foundation Center for Chemistry at the Space-Time Limit (CHE-0802913).

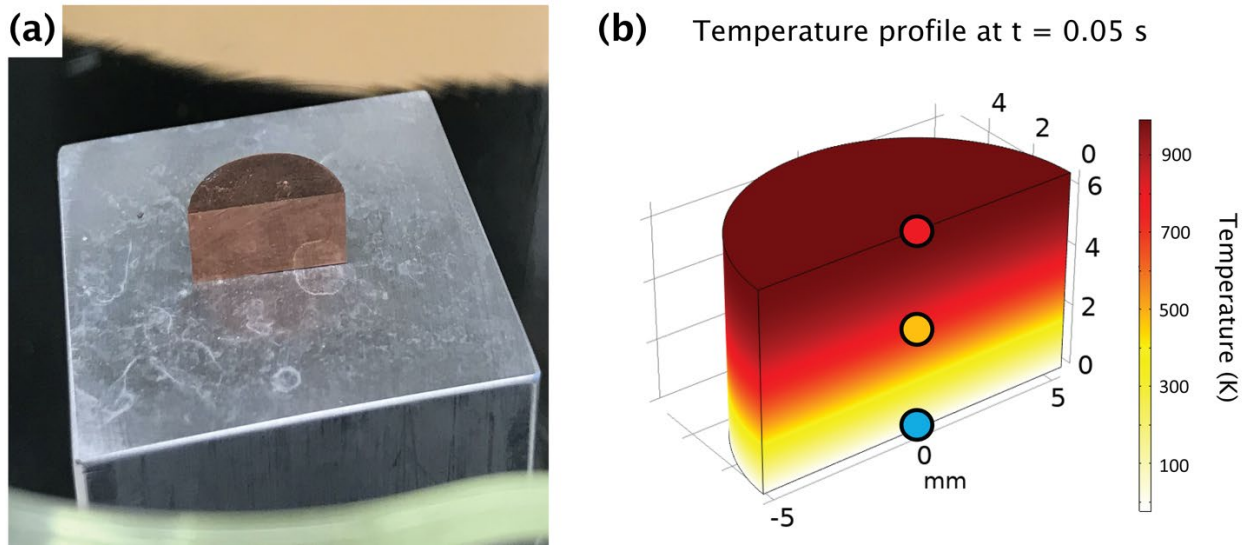

**Figure 1.** (a) Bulk nanocrystalline samples are exposed to variable cooling rates by placing on an Al cooling block in liquid nitrogen, following annealing at 950 °C for 10 min to re-form amorphous intergranular films at grain boundaries. (b) Temperature profile throughout the bulk sample at t = 0.05 s after contacting the cooling block, simulated using COMSOL Multiphysics software. The blue, orange, and red circles represent the fast, medium, and slow quenched locations on the sample, respectively.

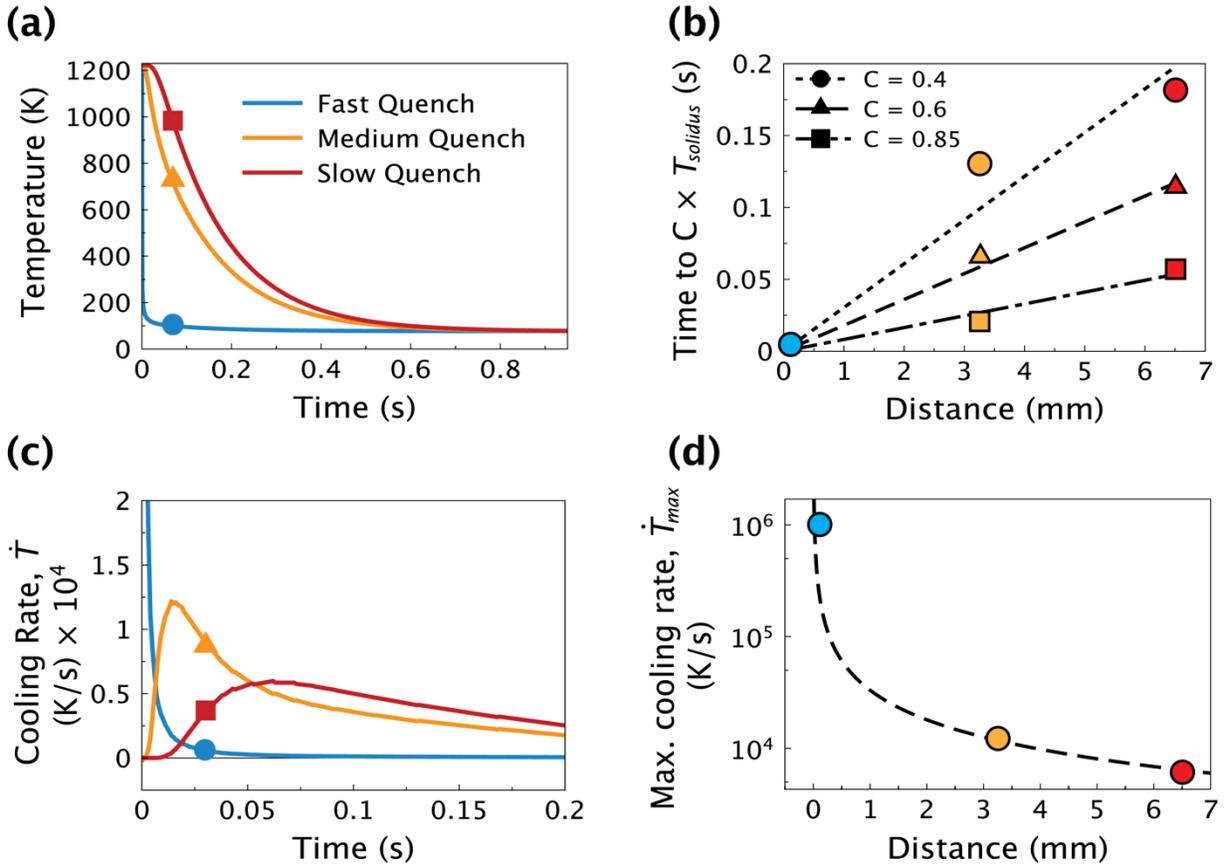

**Figure 2.** (a) Plot of temperature versus time at the fast, medium, and slow quench regions of the bulk samples, as predicted by a simulation of the quenching procedure. (b) Plots of the time for each of the three sample positions to reach a temperature of $0.4T_m$, $0.6T_m$, and $0.85T_m$. $0.85T_m$ represents the highest temperature at which an amorphous-to-ordered complexion transition could potentially occur upon cooling, while $0.4T_m$ represents a possible temperature below which complexion transitions would be restricted due to limited diffusion and atomic rearrangement. (c) Cooling rate as a function of time at the fast, medium, and slow quenched regions of the samples. (d) Maximum cooling rates at the fast, medium, and slow quenched regions as a function of distance from the surface in direct contact with the Al heat sink.

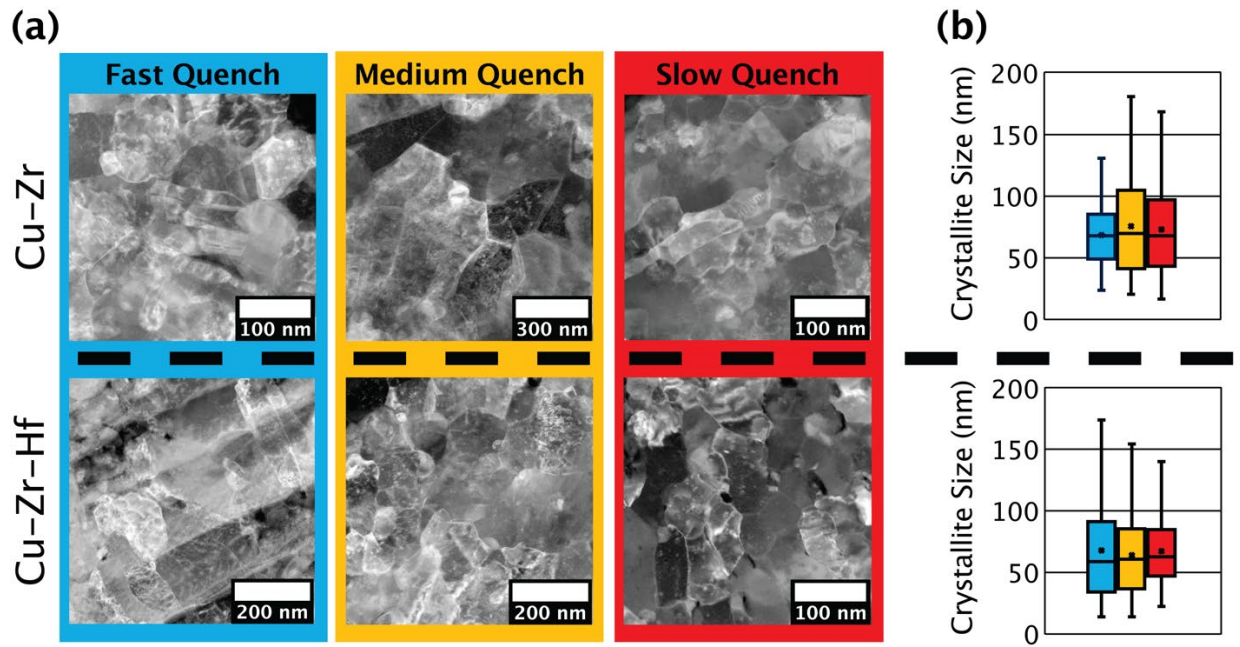

**Figure 3.** (a) HAADF images of the microstructures in the fast, medium, and slow quenched regions for the Cu-Zr and Cu-Zr-Hf samples. (b) Box and whisker plots of crystallite sizes from the three regions in both samples show nearly identical crystallite size distributions in all regions. Dots within the box represent the mean crystallite size for each region, while horizontal lines represent the median crystallite size.

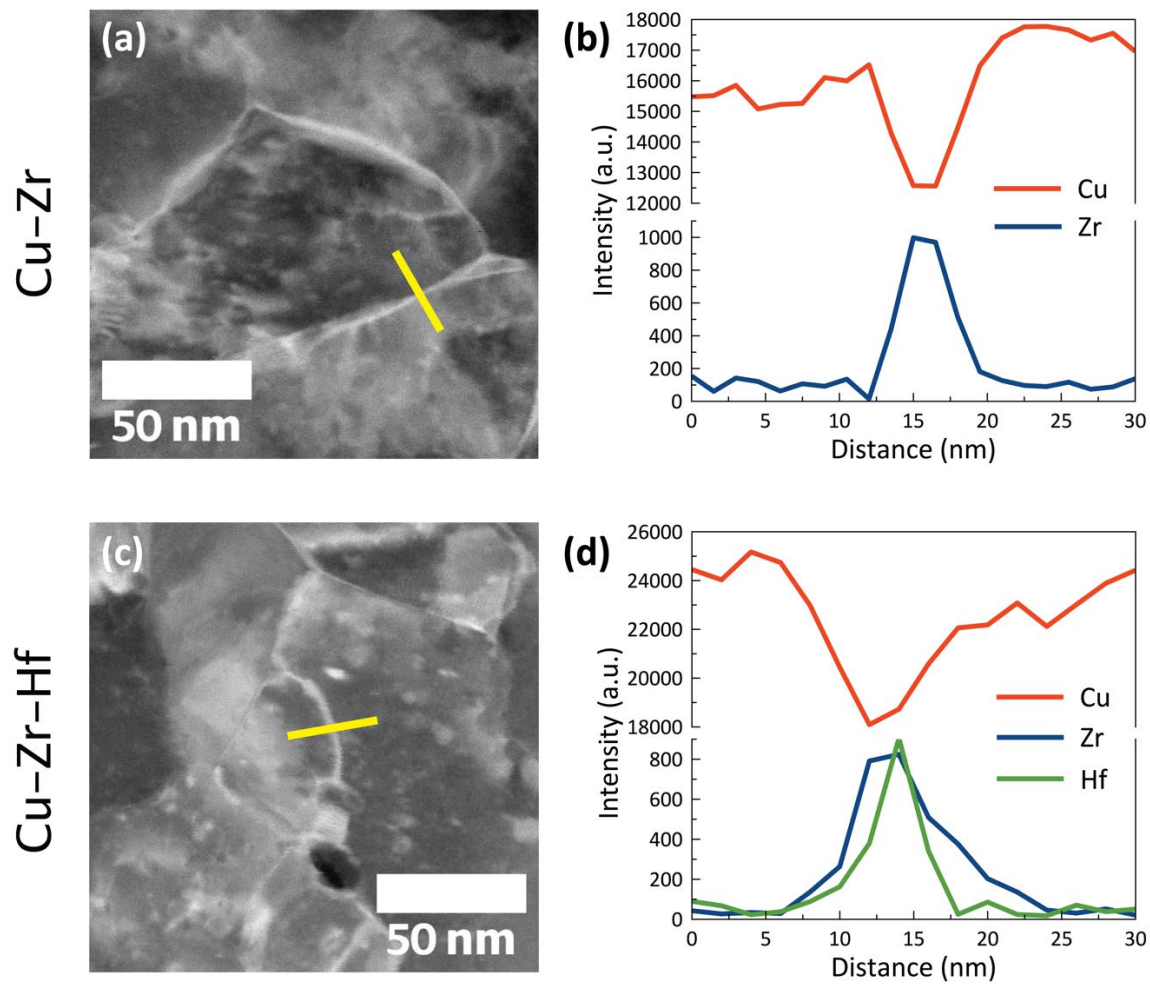

**Figure 4.** HAADF STEM micrographs and STEM-EDS line scans of grain boundary regions from (a, b) the Cu-Zr alloy and (c, d) the Cu-Zr-Hf alloy. In both cases, the dopant concentration is elevated at the grain boundary region, denoting strong segregation.

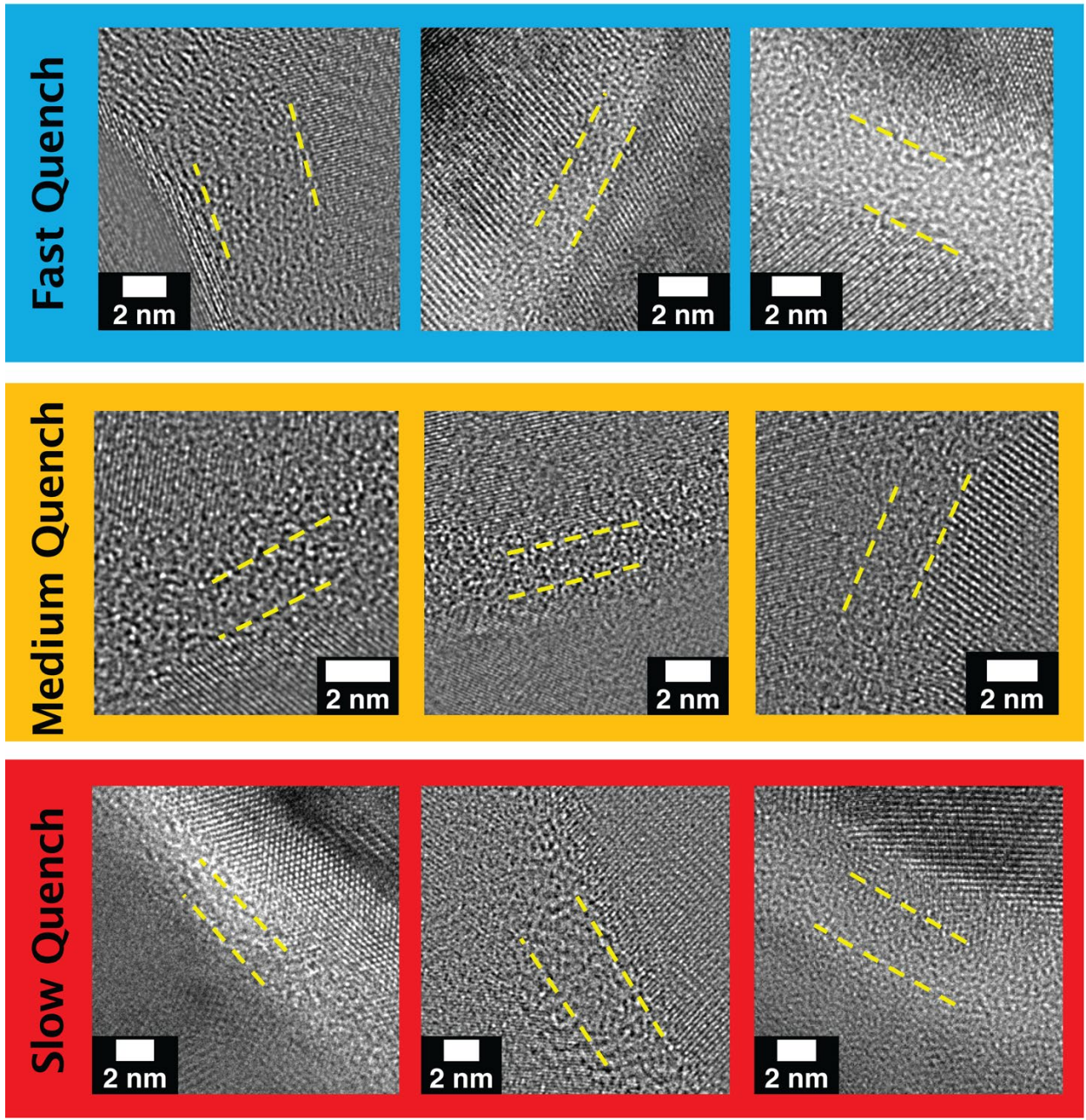

**Figure 5.** HRTEM micrographs of amorphous complexions found in the fast, medium, and slow quench regions of the Cu-Zr sample. Yellow dotted lines denote the edges of the amorphous complexion between abutting grains.

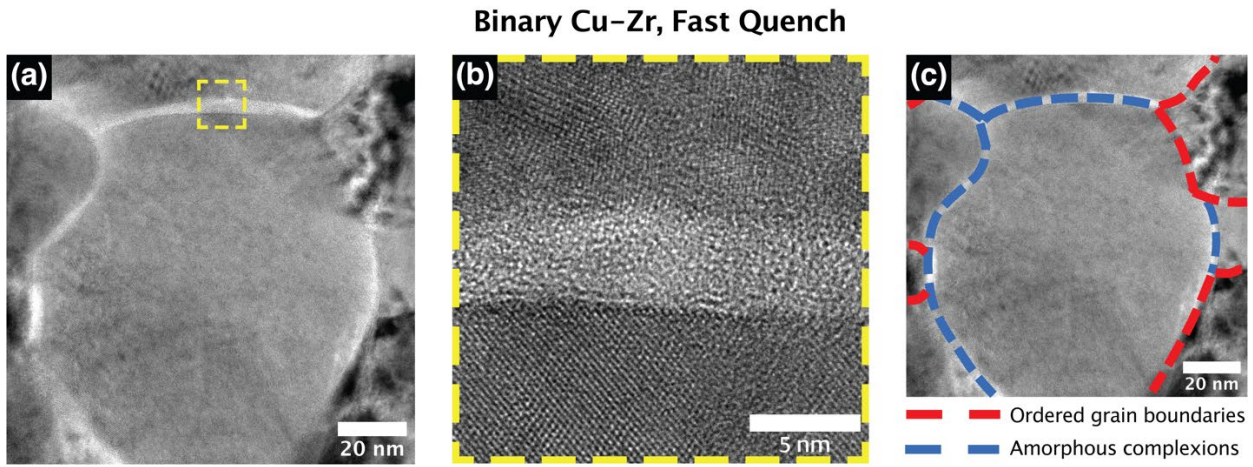

**Figure 6.** (a) TEM micrograph of a grain found in the fast quench region of the Cu-Zr sample that is predominantly surrounded by amorphous grain boundary complexions. (b) HRTEM image of the amorphous complexion along the top of the grain shown in (a). (c) Ordered grain boundaries and amorphous complexions are indicated by red and blue dotted lines, respectively.

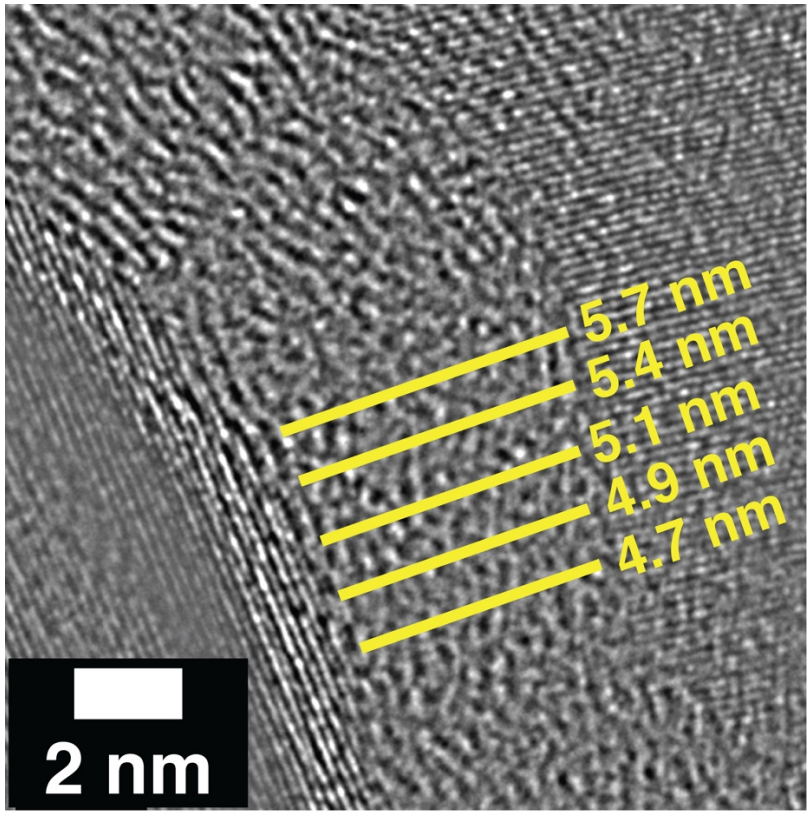

**Figure 7.** HRTEM image of an amorphous complexion in Cu-Zr-Hf, showing that small thickness variations can be observed along the grain boundary due to grain boundary curvature. For consistency, the thinnest region was chosen for each measurement.

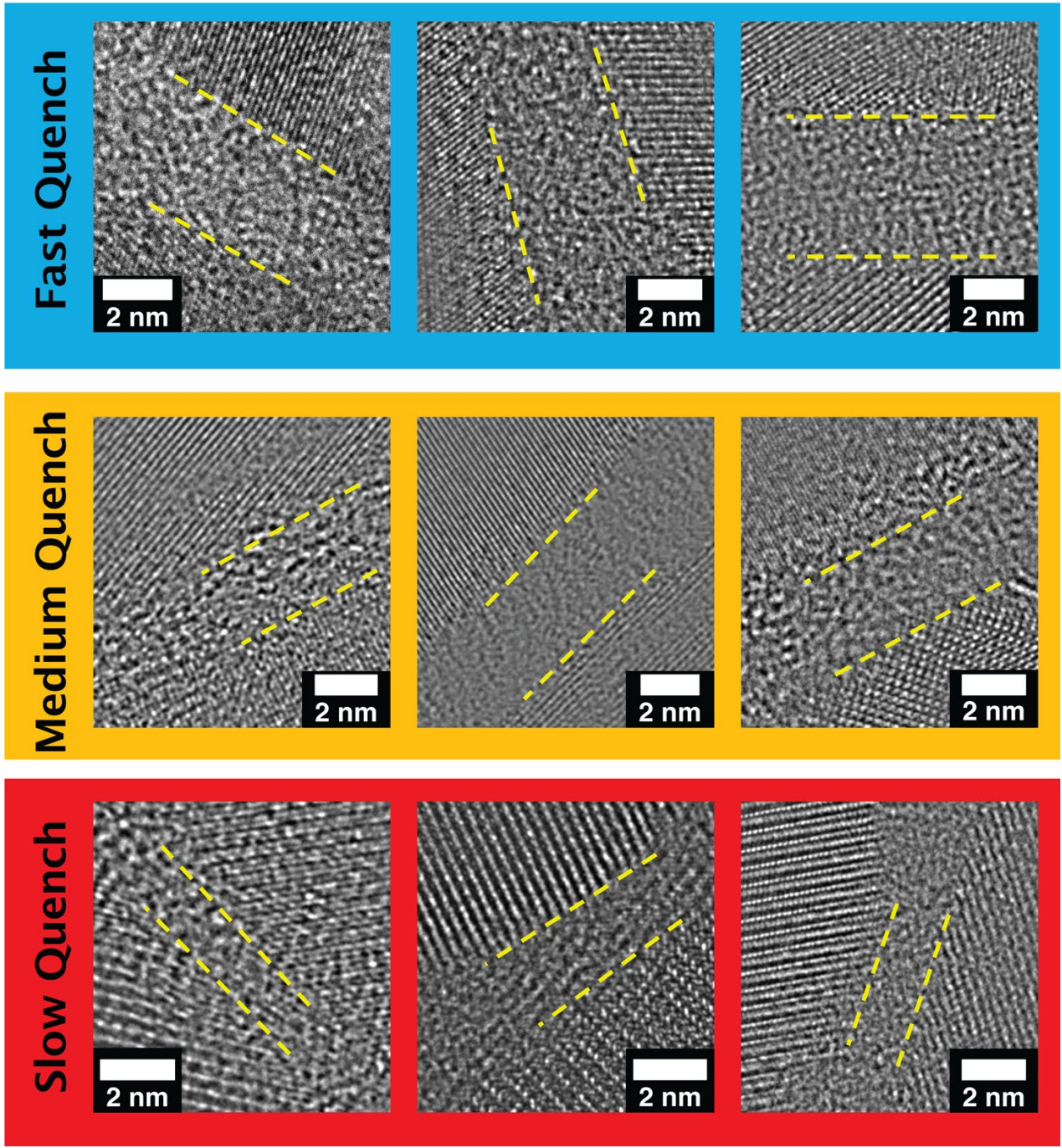

**Figure 8.** HRTEM micrographs of amorphous complexions found in the fast, medium, and slow quench regions of the Cu-Zr-Hf sample.

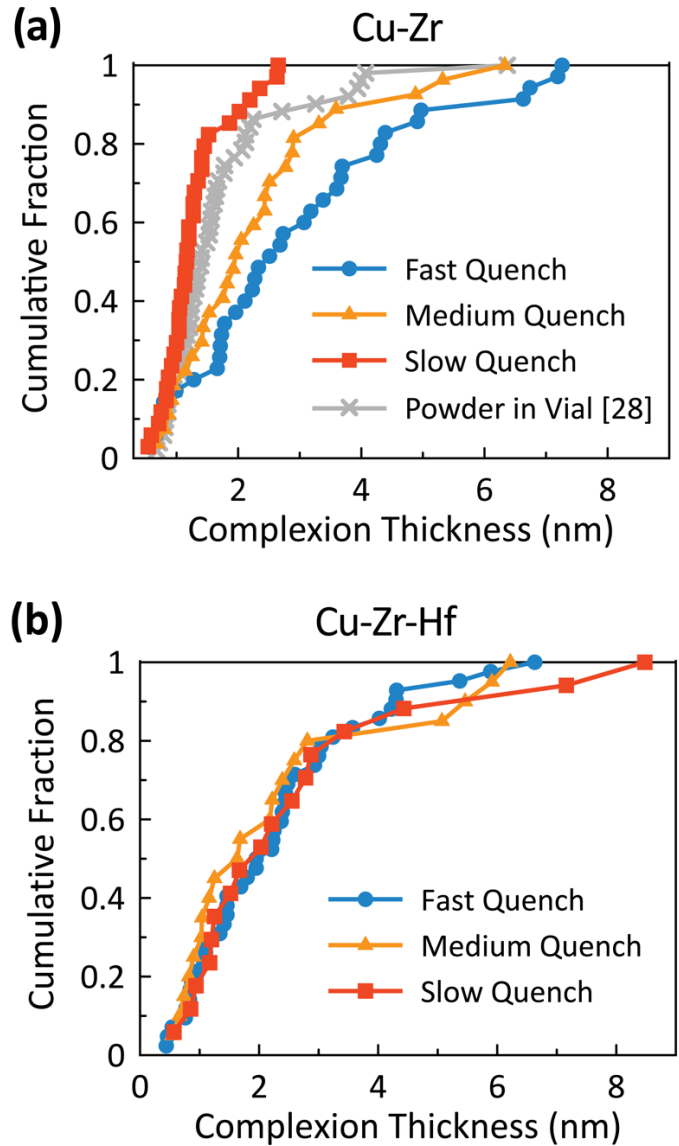

**Figure 9.** (a) Cumulative distribution function of complexion thickness populations in the Cu-Zr sample at the fast, medium, and slow quench regions, as well as from a Cu-4Zr powder sample quenched in a quartz tube investigated in Ref. [28]. (b) Cumulative distribution function of complexion thickness populations in the Cu-Zr-Hf sample at the fast, medium, and slow quench regions, indicating that the average complexion thicknesses in this sample are independent of cooling rate.

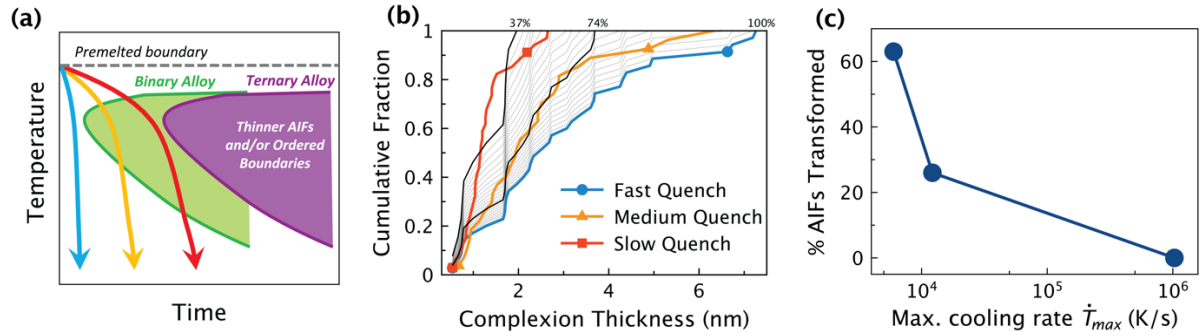

**Figure 10.** (a) Schematic of a time-temperature-transformation (TTT) diagram for the binary Cu-Zr and ternary Cu-Zr-Hf alloys. Blue, orange, and red cooling paths correspond to the fast, medium, and slow quench regions, respectively. The complexion transition region for the ternary alloy is shifted to the right, indicating that AIFs formed in this alloy are more stable against transitions to thinner amorphous films or ordered grain boundaries than those in the binary alloy. (b) Cumulative distribution function of complexion thicknesses in Cu-Zr, where the dashed lines represent subsets of the data from the fast quench sample. The distribution subsets denoted by solid black lines aligned with measurements of the medium and slow quench curves and represent the bottom 74% and 37% of the fast quench data, indicating that 26% and 63% of the amorphous boundaries in these regions have transformed, respectively. (c) Plot of the percentage of AIFs which have transformed to thinner films or ordered boundaries as a function of local maximum cooling rate.

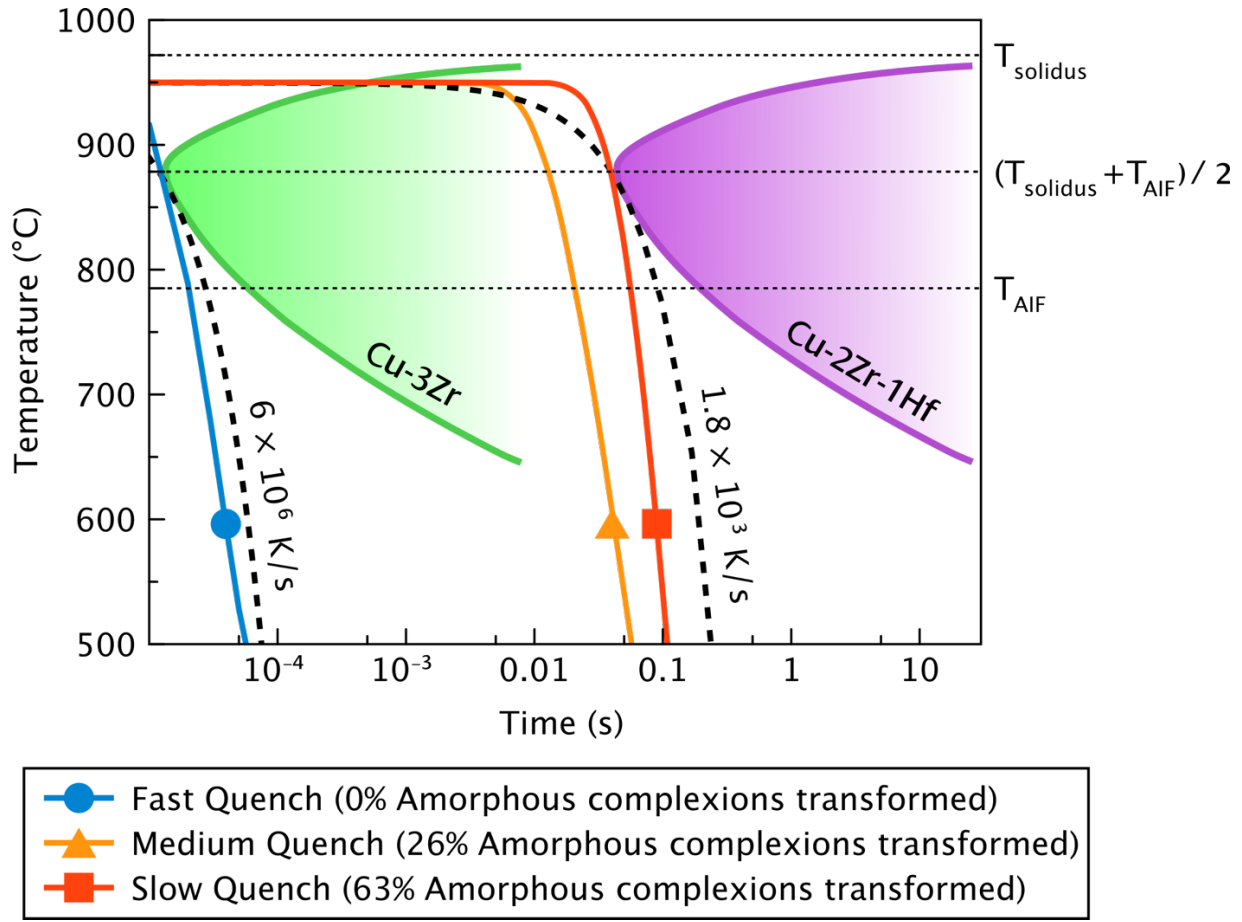

**Figure 11.** TTT curves for Cu-Zr and Cu-Zr-Hf, with plotted cooling paths for the fast, medium, and slow quench regions obtained using COMSOL. $T_{nose}$ for both transition regions is set as the average of the solidus temperature, $T_{solidus}$, and the highest temperature at which amorphous complexions begin to form upon heating, $T_{AIF}$. Estimates of the critical cooling rates necessary to avoid complexion transitions for the binary and ternary alloys are found to be $6 \times 10^6$ K/s and $1.8 \times 10^3$ K/s, respectively, corresponding to a critical cooling rate for the ternary alloy that is roughly 3000 times slower than that for the binary alloy.

**Table 1.** XRD results for both as-milled powders and consolidated bulk samples of Cu-Zr and Cu-Zr-Hf alloys. Crystallite sizes for both the FCC and carbide phases are shown, as well as the volume percent of each phase present. All samples are primarily comprised of the FCC phase and all crystallites are in the nanocrystalline size regime.

|  |  | Crystallite Size (nm) | | Volume Percent (%) | |
|---|---|---|---|---|---|
|  |  | FCC phase | Carbide Phases | FCC phase | Carbide Phases |
| **Cu-3Zr** | As-Milled Powder | 25 | 22 | 99.14 | 0.86 |
|  | Bulk Sample | 59 | 23 | 99.03 | 0.97 |
| **Cu-2Zr-1Hf** | As-Milled Powder | 29 | 27 | 99.09 | 0.91 |
|  | Bulk Sample | 62 | 33 | 98.96 | 1.04 |

**Table 2.** Material properties, sample dimensions, and boundary conditions used for the COMSOL simulation of the quenching process.

**Sample Properties**

|  |  |
|---|---|
| Thermal Conductivity of Cu | 400 W/(m·K) |
| Initial sample temperature | 1223.15 K |
| Cylinder height | 6.47 mm |
| Cylinder radius | 5.34 mm |

**Boundary Conditions**

|  |  |
|---|---|
| Cylinder walls | External natural convection, vertical wall |
| Cylinder top | External natural convection, horizontal plate, upside |
| Temperature at bottom of cylinder | 77 K |
| External temperature | 293.15 K |